\documentstyle[sprocl]{article}

\begin{document}

\title{Sign convention of residues in QCD sum rules}
\author{Seungho Choe}
\address{Department of Physics, Yonsei University,
 Seoul 120--749, Korea}
\maketitle
\abstracts
{ We show that signs of pole residues
  $\lambda_N, \lambda_\Lambda,
\lambda_\Sigma, \lambda_\Xi$
  for ${1\over2}^+$ octet baryons
are identical in the QCD sum rule approach.
To do this we compare signs of
meson-baryon coupling constants $ g_{KN \Lambda}, g_{KN \Sigma}$,
$g_{\pi \Lambda \Sigma}$ and $g_{K\Lambda \Xi}$ each other.}

\section{Introduction}

The method of QCD sum rules has proved to be a very powerful
tool to extract information about hadron properties
\cite{SVZ79,RRY85,Narison89}.
QCD sum rule is based on a study of the following
correlation function (or correlator) of interpolating fields
$j_\Gamma$(x) which are built from quark fields and
have the quantum
numbers of hadrons of interest:
\begin{eqnarray}
\Pi (q^2) = i \int d^4x ~e^{iqx}
    \langle 0| T ( j_\Gamma (x) \bar{j}_\Gamma (0) ) |0\rangle ,
\end{eqnarray}
where the state $|0\rangle$ is the physical nonperturbative vacuum,
and T is the time ordering operator.
For example, for an interpolating field of the proton
one can choose \cite{Ioffe81,RRY85}
\begin{eqnarray}
\eta_N(x) =
\epsilon_{abc}\left[u_a(x)^T C\gamma_\mu u_b(x)\right]
\gamma_5\gamma^\mu d_c(x) ,
\label{ioffe}
\end{eqnarray}
where $u$, $d$ are the up and down quark fields, and $a,b$ and $c$ are
color indices.
 $T$ denotes the transpose in Dirac space, and $C$ is the charge
conjugation matrix.
The lowest-energy contribution to the spectral function is from the
nucleon pole. Its contribution can be constructed from the matrix element
\begin{eqnarray}
\langle0| \eta_N (0) |N(q)\rangle  = \lambda_N u(q) ,
\label{lambda}
\end{eqnarray}
where $|N(q)\rangle$ is a one-nucleon state with four-momentum $q^\mu$
and $u(q)$ is a Dirac spinor for the nucleon.
$\lambda_N$ is the coupling strength which measures the ability
of the interpolating field $\eta_N$ to excite the nucleon from
the QCD vacuum.
In fact, the above interpolating field couples to the negative parity
nucleons also.
Recently, the techniques to study negative parity baryons are given in
Ref.\cite{JKO96}
and Ref.\cite{LK97},
respectively.
In general, the interpolating fields for octet baryons are expressed as
a combination of two fields such as \cite{EPT83}
\begin{eqnarray}
{\cal O}_1^{udu} &=&
\epsilon_{abc}\left[u_a(x)^T C d_b(x)\right]\gamma_5 u_c(x) ,
\nonumber\\
{\cal O}_2^{udu} &=&
\epsilon_{abc}\left[u_a(x)^T C \gamma_5 d_b(x)\right] u_c(x) .
\end{eqnarray}
For example, in the case of the nucleon
\begin{eqnarray}
{\cal O}_N = {\cal O}_1^{udu} + t \cdot{\cal O}_2^{udu} ,
\end{eqnarray}
where t is a mixing parameter. When t = -- 1,
it becomes Ioffe's choice (Eq.(\ref{ioffe})).
Similarly, one can define $\lambda_B$ for another baryons
as in Eq.(\ref{lambda}).
The $\lambda_B$ is an experimentally unknown parameter.
But, in the case of the proton it is related to the proton decay amplitude
\cite{Ioffe81,BIK81,DG82,Belyaev83,BEHS84,CZ84,HIIY86,Gavela89},
or it corresponds to a size of
the proton in bag models \cite{PM84}. Some other interpretations are found
in Ref.\cite{SSV94,Leinweber94}.
In the QCD sum rule approach, we take this $\lambda_B$ from usual
baryon sum rules, but its form is $\lambda_B^2$.
So we do not know the sign.

In the followings we propose how to determine signs of pole residues
$\lambda_N, \lambda_\Lambda, \lambda_\Sigma$ and $\lambda_\Xi$
for ${1\over2}^+$ octet baryons.
To do this we calculate several meson-baryon
coupling constants and compare their signs, each other.

\section{Signs of residues $\lambda_N, \lambda_\Lambda, \lambda_\Sigma,
\lambda_\Xi$}

Recently, using a 3-point correlation function in the QCD sum rule method
 we have calculated the most relevant
coupling constants in kaon production processes:
i.e., $g_{KN \Lambda}$ and $g_{KN \Sigma}$ \cite{CCL96}
(A recent status on these couplings is given in Ref.\cite{BMK97}).
Some other model calculations on these couplings
are found in Ref.\cite{WJC90,GRS92}.

As emphasized in Ref.\cite{CCL96}, we can not predict signs of
these coupling constants within our approach.
The reason is that we don't know signs of the residues $\lambda_N$,
$\lambda_\Lambda$, $\lambda_\Sigma$.
Our results are as follows:
\begin{eqnarray}
- g_{KN\Lambda} &\simeq& {+ \over \lambda_N \lambda_\Lambda} ,
\nonumber\\*
- g_{KN\Sigma}  &\simeq& {- \over \lambda_N \lambda_\Sigma} ,
\end{eqnarray}
where + and -- in the righthand sides mean that the signs of numerator
are + and -- , respectively.
Contributions from higher order corrections and
higher dimensional operators, and the transition term are
usually small. Thus, they can not change the signs of these couplings.
Therefore, comparing with those of de Swart's \cite{AS90}
\begin{eqnarray}
\lambda_N \lambda_\Lambda &\simeq& + sign ,
\nonumber\\*
\lambda_N \lambda_\Sigma &\simeq& + sign .
\label{sign1}
\end{eqnarray}
This means that the signs of $\lambda_N$, $\lambda_\Lambda$ and
$\lambda_\Sigma$ are the same.
In principle, we can change the magnitudes of these couplings with varying
the mixing parameter t.
However, an optimal value of the mixing parameter t which gives
reliable baryon masses is very similar to that of
Ioffe's choice \cite{EPT83,Leinweber97}.
Thus, we concentrate on Ioffe's interpolating fields in our calculations.
There is the other convention \cite{Dumbrajs83}
which gives -- to both $g_{KN\Lambda}$ and $g_{KN\Sigma}$.
But, in this convention our calculation becomes
\begin{eqnarray}
g_{KN\Sigma}  \simeq {- \over \lambda_N \lambda_\Sigma} ,
\end{eqnarray}
and $\lambda_N \lambda_\Sigma$ is + again.
In the followings we present calculation of
$g_{\pi \Lambda \Sigma}$ and $g_{K\Lambda \Xi}$ using the same approach
\cite{CCL96}. According to de Swart's convention the signs of
$g_{\pi \Lambda \Sigma}$ and $g_{K\Lambda \Xi}$ are + and --, respectively.

In the case of $g_{\pi \Lambda \Sigma}$ the sum rule, after
Borel transformation, becomes
\begin{eqnarray}
\lambda_\Lambda \lambda_\Sigma \frac{M_B}{M_\Sigma^2-M_\Lambda^2}
\left(e^{-M_\Lambda^2/M^2} - e^{-M_\Sigma^2/M^2}\right)
g_{\pi \Lambda \Sigma} \frac{f_\pi m_\pi^2}{\sqrt{2} m_q} =
\nonumber\\
- ~{2\over\sqrt{3}} \left(\frac{7}{12\pi^2} M^4
+ \frac{m_s^2}{4\pi^2} M^2
-~m_s \langle \bar{s} s \rangle \right)\langle \bar{q}q \rangle ,
\label{eq:Borel}
\end{eqnarray}
and the coupling constant has the form
\begin{eqnarray}
g_{\pi \Lambda \Sigma} \simeq {+\over \lambda_\Lambda \lambda_\Sigma} .
\end{eqnarray}
It can be a consistency check of Eq.(\ref{sign1}).
We obtain
\begin{eqnarray}
g_{\pi \Lambda \Sigma} \simeq 7.53
\end{eqnarray}
for $\langle\bar{q}q\rangle$ = -- (0.230 GeV)$^3$ and
$\langle (\alpha_s/ \pi) G^2 \rangle$ = (0.340 GeV)$^4$.
A similar result was given in Ref. \cite{BK96} which used the 2-point
correlation function.

Next, in the case of $g_{K\Sigma \Xi}$
the final expression is
\begin{eqnarray}
\lambda_\Sigma \lambda_\Xi \frac{M_B}{M_\Xi^2-M_\Sigma^2}
\left(e^{-M_\Sigma^2/M^2} - e^{-M_\Xi^2/M^2}\right)
\sqrt{2}g_{K\Sigma \Xi} \frac{f_K m_K^2}{2 m_q} =
\nonumber\\
+ \left(\frac{9}{10\pi^2} M^4
+ \frac{7m_s^2}{5\pi^2} M^2
- {6\over 5} ~m_s \langle\bar{s}s\rangle \right) \langle\bar{q}q\rangle .
\end{eqnarray}
The coupling constant has the following form:
\begin{eqnarray}
g_{K\Sigma \Xi} \simeq {-\over \lambda_\Sigma \lambda_\Xi} ,
\end{eqnarray}
and the value of $g_{K\Sigma \Xi}$ is
\begin{eqnarray}
g_{K\Sigma \Xi} = -~7.02
\end{eqnarray}
for $\langle\bar{q}q\rangle$ = -- (0.230 GeV)$^3$ and $f_k$ = 160 MeV.
Therefore, the signs of $\lambda_\Sigma$ and $\lambda_\Xi$ are identical
comparing with de Swart's convention.
It means that the signs of residues for all octet baryons
$\lambda_N$, $\lambda_\Lambda$, $\lambda_\Sigma$ and $\lambda_\Xi$
are the same.

\section{Discussion}

In the previous section we mentioned two conventions for kaon-hyperon-nucleon
coupling constants.
As emphasized in Ref. \cite{AS90} both conventions lead to the same
result for the only physically meaningful sign, $g_{KN\Lambda}$
and $g_{KN\Sigma} \cdot \mu(\Sigma^\circ \Lambda)$.
Here, $\mu(\Sigma^\circ \Lambda)$ is the $\Sigma^\circ - \Lambda$
transition moment.
According to the convention of de Swart, this moment is given by
\begin{eqnarray}
\mu(\Sigma^\circ \Lambda) = -{\sqrt{3}\over 2} \mu_n \simeq + sign ,
\end{eqnarray}
where $\mu_n$ is the neutron magnetic moment.
Although an absolute value of the transition moment can be
measured \cite{PRD},
one can check our result for the sign convention by calculating the moment
in the QCD sum rule approach.
Until now, however, there are no works which calculate the transition moment
from QCD sum rules directly. Only magnetic moments for baryons have been
obtained by QCD sum rules \cite{BY83,IS83,IS84,CPW}
and the transition moment has been obtained from those moments \cite{IS83}.

In conclusion, the relative signs of residues for ${1\over2}^+$
octet baryons are
presented. Their signs are identical in the QCD sum rule approach.
But, we still don't know whether they are + or --.

\section*{Acknowledgements}

The author thanks the organizers, especially Prof. D.-P. Min
and Prof. C.-R. Ji for their effort to make NuSS'97 successful.
He also thanks Prof. Su H. Lee for valuable discussions.
This work is supported in part by KOSEF
through CTP at Seoul National University.


\end{document}